Regulatory mechanisms controlling store-operated calcium entry.

Goutham Kodakandla[1*], Askar M. Akimzhanov[2], Darren Boehning[1*]

[1] Department of Biomedical Sciences, Cooper Medical School of Rowan University, Camden, NJ, USA, 08103

[2] Department of Biochemistry and Molecular Biology, McGovern Medical School, Houston, Texas, USA, 77030

[*]For correspondence, GK (kodaka78@rowan.edu) or DB (boehning@rowan.edu)

**Abstract**

Calcium influx through plasma membrane ion channels is crucial for many events in cellular physiology. Cell surface stimuli lead to the production of inositol 1,4,5-trisphosphate ($IP_3$), which binds to $IP_3$ receptors ($IP_3R$) in the endoplasmic reticulum (ER) to release calcium pools from the ER lumen. This leads to the depletion of ER calcium pools, which has been termed store depletion. Store depletion leads to the dissociation of calcium ions from the EF-hand motif of the ER calcium sensor <u>S</u>tromal <u>I</u>nteraction <u>M</u>olecule 1 (STIM1). This leads to a conformational change in STIM1, which helps it to interact with the plasma membrane (PM) at ER:PM junctions. At these ER:PM junctions, STIM1 binds to and activates a calcium channel known as Orai1 to form calcium-release activated calcium (CRAC) channels. Activation of Orai1 leads to calcium influx, known as store-operated calcium entry (SOCE). In addition to Orai1 and STIM1, the homologs of Orai1 and STIM1, such as Orai2/3 and STIM2, also play a crucial role in calcium homeostasis. The influx of calcium through the Orai channel activates a calcium current that has been termed the CRAC current. CRAC channels form multimers and cluster together in large macromolecular assemblies termed "puncta". How CRAC channels form puncta has been contentious since their discovery. In this review, we will outline the history of SOCE, the molecular players involved in this process, as well as the models that have been proposed to explain this critical mechanism in cellular physiology.

Keywords: calcium, orai1, stim1, dhhc21, s-acylation, store-operated calcium entry, immune diseases.

**Introduction:**

Calcium is a crucial secondary messenger that serves a multitude of functions ranging from subcellular signaling to organ system-level changes. It serves many roles in biological processes, including growth, disease, and death (Clapham, 2007, Carafoli, 2002, Berridge et al., 2000). Calcium levels in the cytosol are maintained at ~100nM in many ways, including calcium pumps and exchangers on the plasma membrane and various organelles. The extracellular calcium concentration is 1-1.5 mM (Clapham, 2007). The calcium ion gradient across the plasma membrane (PM) is maintained by calcium efflux pumps and ion channels, in addition to calcium reuptake mechanisms in ER. Sarco-endoplasmic reticulum calcium ATPase (SERCA) pumps use an active transport mechanism to move calcium across the concentration gradient from the cytosol into the ER lumen (Hasselbach and Makinose, 1961, Ebashi and Lipmann, 1962). Upon cellular stimulation by agonist actions at plasma membrane receptors, a multitude of signaling cascades starts that lead to the production of inositol 1,4,5-trisphosphate ($IP_3$) due to phospholipase-C-mediated cleavage of membrane phosphatidylinositol 4,5-bisphosphate ($PIP_2$) (Rhee, 2001). Once produced, $IP_3$ binds to and activates the $IP_3$ receptors ($IP_3Rs$) on the ER membrane to promote calcium release from the ER lumen into the cytosol. This decrease in calcium levels in the ER leads to intracellular calcium store depletion (Parekh and Penner, 1997). Store depletion leads to the activation of an ER transmembrane protein known as stromal interaction molecule 1 (STIM1), which then binds to Orai1 channels on the plasma membrane to promote store-operated calcium entry (SOCE) (Parekh and Penner, 1997, Parekh and Putney, 2005). The channel formed by Orai1 and STIM1 is known as the calcium release-activated calcium (CRAC) channel. A general overview of SOCE is presented in **Figure 1**.

Store-operated calcium entry is a predominant mechanism for calcium entry from the extracellular milieu to increase cytosolic calcium after store depletion. It also serves to refill the ER calcium stores after IP$_3$-mediated calcium release (Jousset et al., 2007, Negulescu and Machen, 1988). The most important role SOCE plays is a sustained calcium entry that can last from several seconds to up to an hour in some cellular systems (Cheng et al., 2011, Quintana et al., 2011). One specific case where this differential role of calcium entry plays a crucial role is the activation of T cells. The activation of nuclear factor of activated T cells (NFAT), which regulates IL-2 production, a sustained calcium entry is needed, which lasts for several minutes (Hogan et al., 2003). While for activation of NF-KB, calcium oscillations of lower frequency from Orai1 channels are required. Finally, CaMKII, a calcium sensitive kinase, needs calcium oscillations of higher frequency for its activation. Thus, cellular functions regulated by calcium are mediated by changes in both frequency and amplitude (Smedler and Uhlen, 2014). Calcium entry from the plasma membrane is essential for shaping the spatiotemporal aspects of calcium transients, including the amplitude and duration of calcium release events. Calcium entry contributes to many cellular functions such as secretion, gene transcription, and enzyme activation. Efficient SOCE requires the formation of multiple Orai1-STIM1 complexes at defined ER:PM junctions known as "puncta" (Barr et al., 2008, Liou et al., 2007, Penna et al., 2008, Wu et al., 2006). A key difference between CRAC channels and other calcium channels such as voltage or ligand gated ion channels is their ability to conduct calcium currents at resting membrane potentials without ligand binding (Lewis and Cahalan, 1989). This property allows CRAC channels to be gated in non-excitable cells by exclusively by binding to STIM1. This facilitates precise spatiotemporal regulation of cellular calcium signaling by store depletion.

In this review, we look at the historical origins of SOCE and summarize a few landmark studies that established the characteristics of SOCE. We will delve into the proteins that form the CRAC channel complex, such as Orai1 and STIM1, and how these proteins are regulated. In addition to STIM1, STIM2 is another member of ER resident calcium sensor that promotes SOCE. Orai2 and Orai3 are genes in the Orai family with Orai1. Proteins outside these two families such as TRP channels also play a role in SOCE. Finally, we will also review the effect of SOCE-associated regulatory factor (SARAF) and role it plays in SOCE. SARAF promotes calcium dependent inactivation, but how it functions as an inactivator of SOCE is not well known. Our review outlines the key modulators of SOCE. We will conclude with a discussion of very recent work indicating the protein S-acylation is a key mediator of CRAC channel assembly and function.

## **History of SOCE**

The earliest known proposal for SOCE was introduced by the Putney group in 1985 as outlined by their model for receptor-regulated calcium entry, in which they proposed the mechanism for sustained calcium entry from extracellular matrix upon agonist binding to G-protein coupled receptors (Putney, 1986). Termed as the capacitive calcium entry model, they carefully analyzed the calcium release from the ER and entry from extracellular milieu, and the coupling between these two processes. A few years later, Hoth and Penner identified a similar inward rectifying calcium current in mast cells upon depletion of intracellular calcium stores using $IP_3$ (Hoth and Penner, 1992). Zweifach and Lewis then showed store depletion in T cells leads to a similar calcium current using thapsigargin, a SERCA pump inhibitor (Zweifach and Lewis, 1993). Later, studies published by Liou et al, and Roos J et al, identified STIM proteins as sensors of ER luminal calcium (Liou et al., 2005, Zhang et al., 2005). A year later, studies by Feske et al showed how a mutation in Orai1 protein causes immune deficiency by affecting SOCE (Feske et al., 1996). Other studies during this same period also showed Orai1 is crucial for SOCE (Vig et al., 2006b). A genome wide RNAi screen in drosophila S2 cells showed the role of both Orai1 and STIM1 in SOCE. They demonstrated that RNAi of olf186-f, the *drosophila* Orai1 homologue, reduced thapsigargin-evoked SOCE, which was improved by 3-fold upon its overexpression. In addition, co-expression of STIM with olf186-f increased SOCE by approximately 8-fold (Gwack et al., 2006, Yeromin et al., 2006). These seminal studies identified the key regulatory proteins involved in SOCE. What followed is an expansion of studies involving Orai and STIM proteins and their role in CRAC channel formation to promote SOCE.

**IP$_3$-mediated calcium release:**

Cleavage of membrane phospholipids by phospholipase C (PLC) leads to the production of IP$_3$ and diacylglycerol (Rhee, 2001). Liberated IP$_3$ binds to IP$_3$ receptor calcium channels on ER membranes and leads to calcium efflux from the ER lumen (Streb et al., 1983). Depending on the strength and duration of agonist stimulation, IP$_3$R activation leads to a decrease inf calcium levels in the ER lumen which is termed store depletion (Streb et al., 1983, Parekh and Penner, 1997). The role of IP$_3$ in activating calcium release from the endoplasmic reticulum was initially discovered by experiments by the Berridge group in blowfly salivary glands based on the ability of those tissues to respond to hydrolysis of PIP$_2$ (Berridge and Irvine, 1984). Subsequently, the role of IP$_3$ as a secondary messenger was to mobilize internal calcium stores in mammalian cells was demonstrated in saponin-permeabilized hepatocytes (Joseph et al., 1984). Application of IP$_3$ led to a brief rise in intracellular calcium levels followed by a sustained calcium plateau (Joseph et al., 1984). Capacitive calcium entry was the first model proposed to explain the calcium entry observed in cells upon emptying intracellular calcium pools (Putney, 1986). This idea originated from the observation that calcium entry from the extracellular milieu lasted for a long duration after the levels of IP$_3$ returned to the baseline. These observations were made in rat parotid gland upon carbachol (Aub et al., 1982) application, and rabbit ear artery upon application of noradrenaline (Casteels and Droogmans, 1981). Based on the biphasic nature of agonist-induced increase in the cytosolic calcium, it was proposed that the initial increase is a result of calcium release from the ER due to the action of IP$_3$ followed by influx of calcium through channels in the plasma membrane until the levels of calcium in the ER reaches to a significant level that stops the entry (Putney, 1986).

**CRAC channel characterization and models of activation:**

An effort to accurately define the mechanism of SOCE and to distinguish it from other calcium currents led to further research into the channels underlying these currents. CRAC channels conduct calcium currents at a negative membrane potential. The voltage independence of CRAC channel gating was first observed in patch recordings conducted on Jurkat T cells treated with various cell mitogens such as phytohaemagglutinin (PHA), a substance known to activate T-cell signaling pathway (Lewis and Cahalan, 1989). A similar current was also observed in these cells recorded in low extracellular calcium. In a separate set of experiments conducted in mast cells, intracellular dialysis with $IP_3$ and extracellular application of substance P generated a similar low-noise current (1-2pA) in patch-clamp recordings (Matthews et al., 1989, Penner et al., 1988). This current developed in these cells with a concomitant increase of intracellular calcium. The currents observed by both groups showed similar features such as inward rectifying current-voltage relationship, voltage-independent gating, very high calcium selectivity, an extremely low unitary conductance, and extracellular calcium-dependent feedback inhibition.

The discovery of thapsigargin as a potent, selective, and irreversible SERCA pump inhibitor helped in delineating the difference between CRAC currents and other calcium currents (Thastrup et al., 1989). Thapsigargin promotes a slow calcium leak from the ER lumen into the cytosol, and thereby passively depletes ER calcium stores, possibly through the Sec61 translocon on the ER membranes (Van Coppenolle et al., 2004). In addition, the development of calcium sensitive fluorescent dyes made live-calcium imaging feasible without the need for electrophysiological recordings (Grynkiewicz et al., 1985). One established model of isolating SOCE currents using calcium-sensitive dyes involves depletion of ER calcium stores by treatment with thapsigargin in calcium-free buffer. Subsequent treatment of these cells with calcium replete buffer results in calcium entry from extracellular milieu which can easily be monitored using microscopy in live cells. Using this methodology, along with other imaging paradigms, many groups have reported SOCE in both excitable and non-excitable cells. It was discovered that PLC-mediated $IP_3$ production was dispensable for SOCE (Gouy et al., 1990, Sarkadi et al., 1991, Mason et al., 1991). A seminal observation made in parotid acinar cells using thapsigargin to deplete stores garnered evidence that the ER calcium store levels and activate SOCE independently of $IP_3$ production (Takemura et al., 1989). A subsequent set of experiments done on T cells using thapsigargin established the role of ER calcium store depletion and T-cell activation. The term $I_{CRAC}$ was coined by Hoth and Penner after identifying calcium currents in whole cell currents elicited by a number of agents such as ionomycin, $IP_3$, and EGTA (Hoth and Penner, 1992). The crucial experiments conducted by Lewis and colleagues using perforated-patch clamp and thapsigargin in T cells showed the similarity between TCR-mediated calcium currents and CRAC currents (Zweifach and Lewis, 1993, Fanger et al., 1995, Serafini et al., 1995). These studies led to the definitive conclusion that CRAC currents are controlled by intracellular ER calcium stores.

**Hypotheses for CRAC channel activation:**

The research into how ER calcium depletion leads to CRAC channel activation led to three main hypotheses: 1) diffusion of an activating factor released from the ER to the plasma membrane, 2) targeting of active CRAC channels to the PM by membrane fusion, and 3) conformational coupling between a putative ER calcium sensor and a PM calcium channel. Here we will discuss these proposals in detail.

*Calcium influx factor (CIF):* The earliest proposal for a diffusible mediator that is released from the ER into the cytosol and/or the extracellular milieu to activate calcium influx was presented in early 1990s. Application of phytohaemagglutinin (PHA)-treated Jurkat T cell extracts to P388D1 macrophage cells showed a sustained but fluctuating calcium increase. A diluted version of the extract decreased the amplitude and increased the latency of the calcium flux (Randriamampita and Tsien, 1993). Interestingly, NG115-401L cells did not show CIF-induced calcium entry (Thomas and Hanley, 1995). In later studies, it was shown that NG115-401L cells do not express endogenous STIM1 (Zhang and Thomas, 2016). The treatment of putative CIF-containing extracts with alkaline phosphatase neutralized the effect of these extracts. These observations led to a proposal where cellular stimulation leads to the production of a diffusible factor that activates extracellular influx of calcium (Parekh and Penner, 1997, Parekh and Putney, 2005).

*Membrane fusion of active CRAC channels:* It was found that acid extracts from thapsigargin-treated Jurkat cells leads to a chloride current in *Xenopus* oocytes (Thomas and Hanley, 1995). Microinjection of Xenopus oocytes with the Rho GTPase inhibitor clostridium C3 transferase potentiated a calcium entry current termed $I_{SOC}$. In addition, the expression of wild-type or constitutively active Rho inhibited $I_{SOC}$. Interestingly, botulinum neurotoxin A and dominant negative SNAP-25 mutants activated $I_{SOC}$. Treatment of these cells with brefeldin A, an agent that blocks exocytosis by inhibiting protein maturation and exit from Golgi apparatus, has no effect on $I_{SOC}$. Based on these results, Tsien and colleagues proposed the model where SOCE is mediated by exocytosis, leading to CRAC channels being inserted into the plasma membrane (Yao et al., 1999).

*Conformational coupling of an ER calcium sensor and a PM calcium channel:* The $IP_3$ receptor has calcium binding sites in its luminal domain. This calcium binding site on the $IP_3$ receptor was proposed to regulate calcium efflux from the ER into the cytosol (Supattapone et al., 1988, Gill, 1989). Based on these observations, it was proposed that the $IP_3R$ was a sensor for ER calcium levels. Upon depletion of ER calcium stores, a cytosolic domain of the receptor binds to the plasma membrane to promote calcium influx. In this hypothesis, the IP$_3$ receptor is the regulator of calcium homeostasis in the cells, resulting in release from the ER lumen, as well as influx from the extracellular milieu (Irvine, 1990). The role the IP$_3$R plays in this conformational coupling model is analogous to the role of ryanodine receptors and dihydropyridine receptors in muscle cells (Lee et al., 2004).

Identifying the mechanism(s) of CRAC channel activation ultimately required the cloning and characterization of the relevant calcium channel(s) activated by store depletion. In the next section, we will discuss the putative role of transient receptor potential (TRP) channels in mediating SOCE.

**TRP channels as potential CRAC channels:**

Transient receptor potential (TRP) channels are ion channels that show diverse ion selectivity, activation mechanisms, and physiological functions. All TRP channels share some common features such as six transmembrane domains, tetrameric structure, cation selectivity, and sequence homology. The first TRP channels were characterized in *Drosophila* visual transduction mutants (Montell, 2005, Montell, 2001). The *Drosophila trp* and *trpl* mutants have a significant decrease in light-induced calcium influx (Cosens and Manning, 1969, Hardie and Minke, 1992). Combined with other mutants in the PLC pathway which also affected vision, it was hypothesized that *trp/trpl* channels are putative SOCE channels. Cloning and characterization of these channels showed that the protein encoded by the *trp* gene localizes to the eye, contains four N-terminal ankyrin repeats, and a transmembrane topology similar to voltage-gated ion channels. (Montell et al., 1985, Montell and Rubin, 1989). Whole cells currents recorded from Sf9 insect cells showed the channels encoded by *trp* gene are activated upon ER calcium store depletion and are moderately selective for calcium over monovalent cations such as sodium ($P_{Ca}:P_{Na}$ ~ 10:1) (Vaca et al., 1994).

*Drosophila trp channels:* In *Drosophila*, three TRPC members are expressed in the eye that play an important role in phototransduction. TRP, TRPL, and TRPγ are the three proteins work in concert for successful operation of fly vision (Phillips et al., 1992, Xu et al., 2000). TRPL and TRPγ have ~50% sequence identity with TRP in the six TM domains, and only differ in the TRP domain, which contains highly conserved regions known as TRP boxes 1 and 2. Loss-of-function and dominant-negative mutations in *Drosophila* TRPCs have demonstrated the importance of these channels in calcium influx in photoreceptors of *Drosophila* upon light stimulus. The *trpl* mutant flies affect the specificity of different cation influx and a decreased response to light stimulus of long duration. In addition, *trp/trpl* double mutant flies cannot respond to light, showing the importance of these channels. Finally, TRPγ heteromultimerizes with TRPL to form light-regulated channels, and a dominant-negative form of TRPγ suppresses TRPL currents (Reuss et al., 1997, Niemeyer et al., 1996, Leung et al., 2000, Xu et al., 2000).

The *Drosophila* TRP can be activated *in vitro* by blocking the SERCA (sarcoplasmic endoplasmic reticulum calcium ATPase) pumps that maintain the calcium gradient between ER lumen and cytosol (Vaca et al., 1994, Phillips et al., 1992). Thapsigargin irreversibly blocks SERCA pumps and causes a passive ER calcium leak, which will then activate store-operated calcium channels (Thastrup et al., 1990, Thastrup et al., 1989). However, *in vivo* observations do not support the *in vitro* analyses. Thapsigargin-mediated SERCA blockade or $IP_3$ receptor activation does not promote calcium influx or affect phototransduction (Ranganathan et al., 1994, Hardie, 1996, Hardie and Raghu, 1998). Some studies also evaluated the potential effect of diacylglycerol, another product of PLC hydrolysis of $PIP_2$, on calcium influx and phototransduction. Isolated photoreceptor cells from drosophila wild type and *rdgA* (DAG kinase) mutants show constitutive TRP activity. Recently, endocannabinoids produced in Drosophila photoreceptor cells in response to light have been shown to activate TRP channels (Sokabe et al., 2022). In addition, using flies engineered to express genetically encoded ER calcium indicator, the sodium/calcium exchanger has been shown to result in rapid calcium release from the ER upon light exposure (Liu et al., 2020).

*Mammalian TRP channels:* The seven TRPC proteins in mammals are divided into four groups based on sequence homology (Wes et al., 1995, Zhu et al., 1995). Like *Drosophila* TRPC proteins, the mammalian counterparts also include three to four ankyrin repeats, 6 TM domains, and high sequence homology in the TRP box domain in the N-terminus (Wissenbach et al., 1998, Okada et al., 1999, Okada et al., 1998, Philipp et al., 1996, Vannier et al., 1998). Like Drosophila TRPCs, the activation of mammalian TRP channels can be activated in cultured cells through PLC activation (Philipp et al., 1996, Zhu et al., 1996). The specific mediator in the PLC pathway that ultimately activates these channels is still unknown, and hypotheses include activation by $IP_3$, DAG, and calcium. TRPC1-4 proteins have been shown to be activated upon store depletion, either by $IP_3$ or thapsigargin treatment in cultured cells expressing these proteins (Philipp et al., 1996, Zhu et al., 1996, Zitt et al., 1997). Calcium influx factor was proposed to be a factor to activate these channels to promote calcium influx (Randriamampita and Tsien, 1993, Thomas and Hanley, 1995, Smani et al., 2004). Additional hypotheses also included conformational coupling between $IP_3R$ and the TRP channels (Irvine, 1990). Some models also suggested internalization of TRPC1/3/4 channels upon store depletion (Itagaki et al., 2004, Lockwich et al., 2001). Other TRPC proteins, such as TRP6/7, are activated by DAG and show similarities to *Drosophila* TRP channels (Hofmann et al., 1999, Itagaki et al., 2004).

Another complexity in mammalian TRPC channel activation comes from context and cell type-specific activation mechanisms. For example, mouse TRPC2 in vomeronasal neurons has been shown to be activated by DAG, whereas the same protein in mouse sperm has been shown to be activated by calcium release (Lucas et al., 2003, Jungnickel et al., 2001). Whether this differential activation is due to alternate splicing or different multimerization is unknown (Hofmann et al., 2000, Vannier et al., 1999). TRPC3 also shows similar differential activation between DAG and store depletion. Some reports also suggest direct binding between TRPC3 and $IP_3R$ as a mode of activation of these channels (Hofmann et al., 1999, Zhu et al., 1996, Kiselyov et al., 1998)..

*TRP Channels and their role in SOCE:* Of the many roles TRP channels play in cellular physiology, a crucial one is store-operated calcium entry (Brough et al., 2001). TRP channels have low calcium selectivity, which distinguishes them from conventional CRAC channels or voltage-gated calcium channels. As such, TRP channels are high conductance and non-selective cation channels with varying permeability ratios between calcium, potassium, and sodium (Petersen et al., 1995, Xu et al., 1997). The mechanism of TRP channel activation and gating ranges from changes in cytosolic calcium concentration, to membrane depolarization, and external cellular stimuli (Hoth and Penner, 1992). The loop between transmembrane domains 5 and 6 form the channel pore of these TRP channels. Four individual TRP subunits form an active channel complex to promote calcium entry (Venkatachalam and Montell, 2007, Hellmich and Gaudet, 2014). The domain architecture of different TRP family proteins regulates the functions of these proteins. For example, TRPC, TRPV, and TRPA channels have repeats of ankyrin regions in their N-terminus (Hellmich and Gaudet, 2014) and TRPC and TRPM channels have a conserved TRP domain adjacent to the last transmembrane domain (Montell, 2005). TRPM subfamily of TRP channels have a catalytic kinase domain in their C-terminus. These proteins also have a highly conserved TRP box sequence (Glu-Trp-Lys-Phe-Ala-Arg) and proline-rich sequence that regulates signal transduction and gating (Gregorio-Teruel et al., 2014). In addition, coiled-coil domains in C- and N- terminal domains aid in the assembly of some TRP channel subfamilies (Baez-Nieto et al., 2011). These coiled-coil domains also regulate the channel binding to STIM1, the ER calcium sensor, to regulate SOCE (Lee et al., 2014). A C-terminal calmodulin and IP$_3$R binding region in TRPC channels is known to regulate both store-independent, and store-dependent calcium entry (Wedel et al., 2003, Dionisio et al., 2011a).

Support for TRP channels acting as store-operated channels came from the studies done using TRPC1 expression in vitro that showed increased calcium entry after store depletion (Zitt et al., 1997, Zitt et al., 1996, Zhu et al., 1996). One key difference in these experiments was even though these cells showed calcium entry upon store depletion, this current was different from the biophysical properties observed in CRAC currents (I$_{CRAC}$) (Desai et al., 2015, Zitt et al., 1997, Trebak et al., 2003). Experiments later determined that TRPC1 is assembled with the Orai1-STIM1 complex (Ong et al., 2007, Jardin et al., 2008a). Immunofluorescence and confocal microscopy in human salivary gland cells showed colocalization of Orai1, STIM1, and TRPC1 at the plasma membrane upon store depletion (Ong et al., 2007). Activation and gating of TRPC1 is mediated by negatively charged aspartate residues which bind STIM1 at both the SOAR domain as the polybasic domain (Zeng et al., 2008). Examination of how calcium entry observed from Orai1:STIM1 complex upon store depletion differs from TRPC1:STIM1 complex led to many interesting hypotheses. One hypothesis is that the local increase of cytosolic calcium near ER:PM junctions upon store depletion activates cytosolic TRPC1 channels, which then gets embedded into the plasma membrane to be activated by STIM1 (Cheng et al., 2008, Cheng et al., 2011). This model can explain the spatiotemporal differences observed in calcium oscillations in Orai1:STIM1 complexes versus TRPC1:STIM1 complexes. These differences also affect the physiological and pathological outcomes of cytosolic calcium transients. One specific example is the activation of nuclear factor of activated T-cells (NFAT) versus activation of NF-kB. Calcium entry mediated by Orai1 after T-cell stimulation leads to activation of NFAT (Akimzhanov et al., 2010, Feske et al., 2005, McCarl et al., 2009), while TRPC1-dependent calcium entry leads to activation of NFkB (Ong et al., 2012). Interestingly, knockdown of Orai1 in Jurkat cells inhibits both SOC and CRAC currents, with no effect upon knockdown of TRPC1 or TRPC3 (Kim et al., 2009). In addition, calcium entry observed from the Orai1:STIM1:TRPC1 complex leads to insulin release from pancreatic beta cells, platelet activation during blood clotting, and SNARE complex formation for adipocyte differentiation and adiponectin secretion (Galán et al., 2009, Sabourin et al., 2015, Schaar et al., 2019). Furthermore, calcium entry from this complex contributes to prostate and colon cancer cell migration (Perrouin-Verbe et al., 2019, Guéguinou et al., 2016). Finally, calcium entry observed in STIM1-Orai1-TRPC1-TRPC4 complexes plays a pivotal role in right ventricular hypertrophy (Sabourin et al., 2018).

Decades of research has now shown the pivotal role of TRP channels modulating the spatiotemporal aspects of calcium release in many systems. This includes activation by store depletion and other $IP_3$-mediated signaling pathways. However, TRP channels do not re-capitulate the "classic" biophysical properties of store-operated calcium currents originally described T cells, including small unitary conductance and very high calcium selectivity. We will next describe the cloning of Orai1 and STIM1 as mediators of this specific type of SOCE.

**Discovery and cloning of Orai1 and STIM1:**

Feske and colleagues reported abnormalities in T-cell activation in infants born to consanguineous parents that manifested as severe combined immunodeficiency (SCID) (Feske et al., 1996). They found increased CD4+ T cell counts and inability of these T cells to produce IL-2 upon stimulation with phorbol 12-myristate 13-acetate (PMA), concanavalin A, and anti-CD3 stimulation. Exogenous application of IL-2 restored the proliferative deficiencies of these cells. Further analyses into the DNA binding activity of transcription factors showed a lack of NFAT binding to DNA with no difference in AP-1, NF-kB, and Octamer binding proteins. They also found no differences in the expression of NFAT, but the dephosphorylation and nuclear translocation was affected in these T cells (Feske et al., 2000a). In subsequent experiments using PMA+ionomycin and increasing concentrations of extracellular calcium, they found that higher levels of extracellular calcium rescued IL-2 production in these T cells (Feske et al., 2000b). They concluded that dysregulation in upstream signaling events play a role in NFAT binding to DNA elements that ultimately results in SCID (Feske et al., 1996).

During this time, it was discovered that there exist two different calcium mobilization patterns in T cells after stimulation. An immediate transient calcium spike (Berridge, 1993) followed by a sustained calcium influx from extracellular milieu (Clapham, 1995, Putney, 1986). The transient calcium release activates NF-kB and JNK, but NFAT activation needs sustained elevations to promote dephosphorylation by the calcium-dependent phosphatase calcineurin (Dolmetsch et al., 1997, Timmerman et al., 1996, Kiani et al., 2000, Crabtree, 1999). Calcium imaging of T cells obtained from the patients with SCID showed dysregulated calcium entry upon stimulation with anti-CD3, ionomycin, and thapsigargin. More importantly, membrane hyperpolarization with valinomycin did not alter calcium entry abnormalities found in these cells, suggesting the lack of calcium entry is not a result of depolarization (Feske et al., 2001). These intriguing results warranted further research into the calcium entry in T cells upon their activation and the ion channels that promote sustained calcium entry.

Development of RNAi screening as a research tool in early 2000s opened avenues for high throughput identification of proteins that mediate specific signaling pathways (Dykxhoorn and Lieberman, 2005, Downward, 2004). Two important siRNA screens performed at this time identified the proteins involved in mediating $I_{CRAC}$ and pinpointed the possible mechanism of activation (Gwack et al., 2006, Zhang et al., 2006). These two screens were performed in the Drosophila S2 cells, which were known to have a store-operated channel with low conductance and high calcium selectivity similar to $I_{CRAC}$ in mammalian cells (Yeromin et al., 2004). The siRNA screen performed by Feske and colleagues identified known proteins that affect calcium influx in addition to NFAT regulatory proteins (Gwack et al., 2006). When they used this RNAi screen to identify potential regulators of $Ca^{+2}$/calcineurin mediated NFAT activation, they found a kinase that negatively regulated exogenously expressed NFAT. In addition, they also found other candidates in this screen that alter calcium levels in the cytosol such as SERCA, Homer, and STIM. Concurrently, Zhang and colleagues independently performed a similar genome-wide RNAi screen (Zhang et al., 2006). Probing for hits that resulted in an inhibition of calcium influx after store depletion by thapsigargin, they identified 11 transmembrane proteins including STIM. Of note was another four transmembrane protein olf186-F, which showed a reduction in SOCE and CRAC currents. They followed up this experiment with an overexpression paradigm, which showed a three-fold increase in CRAC currents, which was further increased to eight-fold upon co-expression with STIM (Zhang et al., 2006). These two pivotal experiments identified Orai1 and STIM1 as the core proteins that form the CRAC channel.

At the same time these RNAi experiments were being conducted, Feske and colleagues continued their research into understand the T-cell activation abnormalities in SCID infants. Using whole-cell patch-clamp electrophysiology performed on T cells from control and SCID patients, they showed that T cells obtained from some SCID patients lacked SOCE. Furthermore, they also found that exogenous expression of STIM1 in these SCID cells did not rescue SOCE in these cells (Feske et al., 2005). Next, they used a combined approach of *Drosophila* RNAi screening and linkage analysis with a single-nucleotide polymorphism (SNP) array to screen for regulators of SOCE and nuclear import of NFAT. T cells obtained from SCID infants and relatives of these infants were analyzed for SOCE deficits and SNP mapping arrays using their genomic DNA. This, in conjunction with *Drosophila* RNAi screen for NFAT regulators, pointed to a gene olf186-f in Drosophila and to the *TMEM142A* gene on chromosome 12 in humans which they named Orai1 (Feske et al., 2006). Genomic DNA sequencing showed a mutation of an arginine to tryptophan at residue 91 (R91W) on Orai1, identifying the molecular defect in these patients leading to reduced SOCE. Finally, exogenous expression of wild-type Orai1 in T cells from SCID patients restored CRAC channel activity.

STIM1 is mammalian homolog of *Drosophila* STIM, which is essential for CRAC channel activation in human T cells. Roos and colleagues showed in S2, HEK293, and Jurkat cell lines that silencing of STIM1 abrogates SOCE (Roos et al., 2005). Zhang et. al. determined the role of the luminal EF hand motif by expressing calcium binding mutants in Jurkat T cells and analyzing SOCE. The STIM1 EF hand calcium binding mutants 1A3A and 12Q were found to be constitutively active STIM1 mutants (Zhang et al., 2005). At the time, the pore forming Orai1 subunit had not been cloned and it was hypothesized that STIM1 may traffic to the PM upon store depletion to activate a channel or form a channel itself. The critical role of the EF hand in SOCE and puncta formation was independently confirmed by another group in HeLa cells using a D76A mutant (Liou et al., 2005). This discovery, along with the discovery of Orai1 as CRAC channel pore forming subunit, potentiated a barrage of research into these proteins in SOCE.

The discovery of Orai1 as pore-forming subunit of CRAC channels was a direct result of the RNAi approaches in *Drosophila* S2 cells by several groups as outline above. Following the research into calcium deficits observed in SCID human patients, two different groups pursued RNAi screens focused toward decreases in NFAT translocation into nucleus, and they individually found the same four transmembrane plasma membrane protein with intracellular C and N termini olf186-F (Feske et al., 2001, Vig et al., 2006b), and named it CRACM1 (Vig et al., 2006b). The human homolog of this protein was eventually named Orai1. There are three homologs of Orai proteins with high sequence identity, with up to 90% identity in the transmembrane regions. Overexpression of Orai1 and STIM1 in HEK293 cells helped identify the pore forming subunit of CRAC channels. The definitive proof for Orai1 as the pore forming subunit came from experiments that showed the importance of acidic residues in the transmembrane domains of Orai1. Residues E106 and E190 are two important glutamic acid residues that line the pore of Orai1 channel. Mutation of these residues to aspartate and glutamine respectively decreased the calcium influx, increased monovalent cation current, and made the channel cesium permeable (Prakriya et al., 2006). These experiments strongly supported Orai1 as the pore forming unit of CRAC channel.

**Molecular players in SOCE**

The discovery of STIM1 as ER calcium sensor that activates SOCE upon store depletion was a pivotal moment in furthering the field of store-operated calcium entry (Liou et al., 2005). Soon after the discovery of STIM1, multiple research groups discovered a four transmembrane cell surface protein in *Drosophila* and its homologs in mammalian cells which would ultimately be known as Orai (Feske et al., 2006, Zhang et al., 2006, Vig et al., 2006b). Overexpression of Orai1 in T cells isolated from patients with SCID restored the calcium entry deficits in those cells. In addition, co-expression of Orai1 with STIM1 in HEK293 cells showed calcium currents in these cells that matched the biophysical properties of CRAC currents (Mercer et al., 2006). CRAC channel formation is interesting in that is requires the assembly of proteins from two different membranous subcellular compartments to form the active channel complex (Kim and Muallem, 2011, Lioudyno et al., 2008, Ong et al., 2007, Zhang et al., 2005, Zhou et al., 2010a). In this next section, we will discuss how this mechanism of activation was discovered.

**STIM1**

Depletion of ER calcium stores initiates a cascade of events starting with the activation of STIM1 and culminates with calcium entry from the extracellular milieu. STIM1 is a dimer of two single pass type I ER transmembrane proteins. In the N-terminus, there is a canonical and a noncanonical EF hand motifs that play a role in calcium sensing in the ER (Stathopulos et al., 2006, Stathopulos et al., 2009). Immediately adjacent to the EF hand motifs, there is a sterile alpha motif (SAM) domain. On the cytosolic side, there is a calcium-activating domain (CAD) which is also known as STIM-Orai activation region (SOAR), which binds to the Orai1 channel (Yuan et al., 2009). The cytosolic C-terminus also contains three coiled-coil domains (CC1,

CC2, CC3) that help maintaining STIM1 in its inactive conformation when ER calcium stores are full. Near the C-terminus, there is a polybasic domain which binds to the $PIP_2$ in the plasma membrane (Yuan et al., 2009). In addition to these domains, STIM1 also has and inhibitory domain (ID), a proline-serine rich domain (P/S domain), EB1 binding domain (EB) in its cytosolic side. These domains act in conjunction with each other upon store depletion to colocalize with Orai1 to form CRAC channels.

The calcium binding EF hand motifs are required for STIM1 to sense ER calcium store levels. This was confirmed using experiments with mutations in the EF hand motif. The mutations D76A, D78A, and E87A reduce the affinity of this motif to bind to calcium and show constitutive puncta formation as well as CRAC currents in resting cells (Liou et al., 2005, Mercer et al., 2006, Zhang et al., 2005). The EF-SAM domains of STIM1 and STIM2 show similar affinities to calcium *in vitro*, however, structural differences in the two proteins results in dramatically different abilities to oligomerize in response to store depletion (Stathopulos et al., 2006, Zheng et al., 2008).

The first step in the CRAC channel activation after ER calcium store depletion is the conformational change in STIM1. This was elucidated in experiments conducted using the cytosolic fragment composed of amino acids 233-685, which is capable of activating SOCE regardless of store depletion (Huang et al., 2006). Further experiments involving truncation mutants of STIM1 led to identification of STIM1-Orai1 activating region (SOAR/CAD), the stretch of amino acids that can activate SOCE without store depletion. It is, however, interesting to note the difference in the amino acid sequences identified by individual groups ranged from 342-448 (Park et al., 2009), 344-442 (Yuan et al., 2009), and 339-444 (Kawasaki et al., 2009). Despite these differences, the consensus is that the SOAR/CAD region can activate CRAC currents independently of store depletion. Further research into CAD domain also showed that CAD binds both the N- and C-termini of Orai1, but the strength of binding is higher at the C-terminus (Park et al., 2009). This binding between Orai1 and CAD is mediated by a coiled-coil interaction between the CC2 region of STIM1-CAD and the polar residues in the C terminus of Orai1 (Stathopulos et al., 2013).

In resting conditions, STIM1 is diffusely localized throughout the ER, but upon store depletion redistributes into clusters that are termed "puncta" (Park et al., 2009). In resting conditions, STIM1 is in a compact conformation that keeps it from interacting with Orai1. Upon store depletion, a conformational change in STIM1 leads to an extended conformation with its N-terminus reaching toward the plasma membrane where it binds to the C-terminus of Orai1 (Wu et al., 2006, Xu et al., 2006, Zhang et al., 2005). Following store depletion, a complex choreography of intramolecular events happens between several domains of STIM1 that stabilizes STIM1 in its extended conformation to promote binding with Orai1. The ER:PM sites of cells are held in close apposition to promote STIM1 and Orai1 binding to promote CRAC channel formation by a multitude of accessory proteins such a septins, synptogamins, junctate, and others as described above (Wu et al., 2006). The extended conformation of STIM1 was proposed to trap Orai1 into puncta, leading to channel gating to promote SOCE (Wu et al., 2014, Hoover and Lewis, 2011). This is known as the "diffusion-trap" model.

STIM1 exists as a dimer in cells with calcium replete stores in the resting state. This was confirmed using the C-terminal cytosolic fractions *in vitro* which formed dimers in solution. The specific domains involved in dimerization of STIM1 were further resolved using co-immunoprecipitation and fluorescence photobleaching of individual fragments (Covington et al., 2010). The coiled-coiled domain fragments and CAD domain independently can form dimers *in vitro*. The CAD domain dimerization is interesting because STIM1 binding to Orai1 is also mediated by this domain. The CC1 domains alone can form dimers, but they are weak and unstable (Covington et al., 2010, Muik et al., 2009). This supports the conclusion that STIM1 dimerization is a complex process requiring the interaction of multiple domains in the protein.

The CAD domain of STIM1 can independently and constitutively activate Orai1 and promote CRAC currents. Based on a crystal structure of CAD domain, hydrophobic residues form hydrogen bonds between the two monomers to stabilize the dimer (Kawasaki et al., 2009). In addition, interactions between the CC2 and CC3 regions of the STIM1 are purported to stabilize the extended conformation of STIM1. One interesting finding in the structure of CAD domain is the role of the CC2 and CC3 helices in binding to and activating Orai1 channels. In resting inactive state, the two CC2 helices (in a dimer) are in a parallel configuration in a tight hairpin structure with CC3 helices. Upon store depletion, these CC2 helices pivot and twist to an antiparallel orientation, allowing CC3 to extend out and allow for binding with Orai1 (Fahrner et al., 2014, Stathopulos et al., 2013, Stathopulos et al., 2008, Yang et al., 2012). These are highly complex and precise molecular movements in a restricted space between ER and PM within seconds of store depletion.

The EF hand domains of STIM1 residing in the ER lumen also undergo dimerization upon release of calcium from their binding pockets. In resting state, these domains exist as compact monomers bound to calcium (Stathopulos et al., 2006). NMR-resolved calcium-bound STIM1/2 EF-SAM fragments show an α-helical structure with a canonical (cEF) and non-canonical (nEF) EF hand domains. The cEF domain contains a helix-loop-helix structure that binds to calcium and nEF domain does not bind to calcium but stabilizes the cEF through hydrogen bonding (Zheng et al., 2011). Upon calcium release, the EF-SAM domains unfold leading to conformational changes in luminal and cytosolic domains of STIM1. Mutations in the glutamate (E87A) or leucine residues (L195R) individually, and phenylalanine and glycine (F108D & G110D) together, in the EF-hand domain led to puncta formation irrespective of calcium depletion (Stathopulos et al., 2008). This led to a proposal that calcium release from EF-hands leads to STIM1 oligomerization upon store depletion. Live-cell imaging and FRET studies using STIM1 mutants added support to this hypothesis, as they show increased FRET between STIM1 fused with YFP and CFP in RBL cells, which was reverse upon calcium addback (Liou et al., 2007). Furthermore, FRET experiments conducted using Orai1-CFP and STIM1-YFP also show a spatiotemporal correlation between STIM1 oligomerization and Orai1-STIM1 binding (Muik et al., 2008).

**Orai1**

There are three homologs of Orai proteins (Orai1, Orai2, Orai3) in humans, and all three proteins promote calcium entry upon store depletion with varying biophysical properties. They can also form both homo- and hetero-multimers. For example, Orai1 and Orai3 can assemble as heteromultimers to promote store-independent calcium channels that are regulated by arachidonic acid or leukotriene $C_4$ (Gonzalez-Cobos et al., 2013, Mignen et al., 2008a, Mignen et al., 2009, Thompson and Shuttleworth, 2013). These are called ARC channels and LRC channels, respectively.

Orai protein subunits are composed of approximately 300 amino acids. Early evidence from co-immunoprecipitation and FRET experiments suggested that these proteins are oligomers in functional CRAC channels (Vig et al., 2006a, Muik et al., 2008, Navarro-Borelly et al., 2008, Gwack et al., 2007). Experiments using preassembled tandem Orai1 multimers and co-expression with dominant-negative Orai1 mutants showed that Orai1 may form homotetramers in CRAC channels similar to other ion channels (Mignen et al., 2008b). Using photo-bleaching of individual fluorophores in tandem Orai1-STIM1 multimers suggested a similar result of four Orai1 molecules with two STIM1 dimers, which was confirmed using FRET (Ji et al., 2008). These results were replicated in live mammalian cells as well as cellular lysates obtained from lymphocytes (Madl et al., 2010, Penna et al., 2008). However, the X-ray crystallographic structure of Drosophila Orai revealed a hexameric stoichiometry challenging these observations (Hou et al., 2012).

The X-ray and CryoEM structures of the *Drosophila* Orai channel helped resolve the oligomeric status of Orai and revealed the mechanisms for gating and ion permeation (Hou et al., 2012, Hou et al., 2020, Hou et al., 2018). This structure shows Orai1 as a four transmembrane protein with cytosolic C and N termini. The TM1 forms the channel pore with the Orai1 N terminus. TM2 and TM3 shield the pore from the membrane and TM4 and C terminus extend away from the channel pore. The channel pore has a ring of glutamate residues on TM1 (E178 is Drosophila, E106 in humans) that form a highly negative electrostatic region and serves as a selectivity filter. The side chains of these glutamate residues extend into the central pore, where the oxygen atoms of the carboxylic groups are in close proximity (~6 Å). Below the selectivity filter is a region of hydrophobic amino acids followed by positively charged residues that extend into the cytosolic lumen. One interesting observation from the crystal structure is that the C terminal tails of Orai1 helices form anti-parallel helices with each other in the hexamer. These interactions are held together by hydrophobic interactions between leucine residues at 316 and 219 (273 and 276 in humans). Mutations at these residues, such as L273S/D and L276D, lower the coiled-coil probability of the C terminus as well as inhibit STIM1 binding and channel activity (Li et al., 2011, Li et al., 2007, Navarro-Borelly et al., 2008). However, an NMR structure between Orai1 272-292 fragment and 312-387 STIM1 fragment shows that the anti-parrel orientation doesn't change upon binding, but only leads to a small change in the angle or Orai1 helices (Stathopulos et al., 2013). The information obtained from these multitude of experimental approaches revealed key details regarding the gating of CRAC channels, especially how the allosteric modulators regulate Orai1 channel function. Four seminal studies helped us elucidate the structural features regulating gating and ion permeation in the Orai channel (Hou et al., 2020, Hou et al., 2018, Hou et al., 2012). Glutamate residues in the extracellular side of TM1 (E106, E178 in *Drosophila*), as explained earlier in the review, were hypothesized to form a selectivity filter by selectively binding calcium over other divalent cations. Mutation of a valine at position 102 (V174 in *Drosophila*) in the TM1 domain to alanine or serine made the channels constitutively active (McNally et al., 2012). The *Drosophila* Orai structure also shows the hydrophobic valine residues making extensive connections protruding into the ion permeation pathway that presents a de-solvation barrier for ions in closed configuration of the channel (Hou et al., 2012, Dong et al., 2013). Using terbium luminescence and disulfide crosslinking, others have also shown that STIM1 binding to Orai1 leads to a conformational change in the extracellular side near E106 and V102, and that this short segment forms a STIM1-dependent barrier (Gudlur et al., 2014). Based on these observations, V102 was proposed as a hydrophobic gate, which was eventually refuted and F171 is now considered to be the hydrophobic gate (discussed in more detail below). Surprisingly, mutation of glycine and arginine residues in the intracellular side of TM1 region of Orai1, at 98 and 91 respectively, resulted in constitutive activation of Orai1 channels. This led to a speculation whether there is another gate in the in the cytosolic side of the channel (Zhang et al., 2011). Based on these observations, R91 was thought to be the physical gate at the intracellular side that leads to dilation of the helices upon STIM1 binding. G98 was suggested to serve as a hinge, upon which the N terminus can rotate to allow for calcium entry. In addition, a phenyl alanine at 99 is on the opposite side of the pore helix to G98. STIM1 binding has been proposed to evoke a conformational change that exposes G98 while concealing F99 away from the channel pore (Yamashita et al., 2017). The *Drosophila* Orai crystal structure revealed basic residues in the immediate inner cytosolic end of channel pore which led to another hypothesis where these residues stabilize the closed channel through either binding of anions or electrostatic repulsion of cations near this pore (Hou et al., 2012). With the knowledge obtained from structural and functional studies using Orai1 and STIM1 mutants, several hypotheses have been put forward to explain the mechanism of activation of Orai1 by STIM1. STIM1 has been proposed to activate Orai1 in stepwise manner, first by initial binding to the C terminus for docking and a subsequent weaker binding to the N terminus to initiate conformational changes in the TM pore leading to pore opening (Zheng et al., 2013).

A stretch of conserved acidic amino acids of Orai1 have been shown to confer calcium selectivity. These include residues at E106, E190, D110, D112, and D114 all of which are either in the TM1 or TM1-2 loop. Mutation of aspartate residues to alanine resulted in increased cesium permeability in these channels (Zhou et al., 2010b, Yamashita et al., 2007). In addition, mutation of E106 residue in the channel pore resulted in the same increase in cesium permeability. These mutants also increase the pore diameter as evidenced by increased permeability of methylammonium derivatives (Yamashita et al., 2007). Finally, E106D mutant also increases the fast calcium-dependent inactivation time latency compared to wild-type Orai1 (Yamashita et al., 2007).

The structure of the open conformation of Orai1 was recently resolved at 3.3 Å using cryo-electron microscopy (Hou et al., 2020). As mentioned earlier, the hydrophobic amino acids F99 and V102 (F171 and V174 in human) within the channel pore were thought to function as channel gate (Hou et al., 2012). This structure shows the channel pore lined by acidic residues facing the extracellular entrance by D184, D182, Q180 and E178 followed by hydrophobic residues V174, F171, and L167 lining the middle of the pore, and positively charged K163 and K159 lining the pore on the intracellular side. A comparison of the open and closed structures revealed that F171 is the primary gate, with the side chain rotating into the ion conduction pathway in the closed state (Hou et al., 2020). Amino acids Q180, D182, and D184 on the extracellular surface create a negative electrostatic surface to attract cations near the mouth of the pore with E178 acting as a selectivity filter at the entrance to the ion conduction pathway.

**SARAF**

SARAF (SOCE-associated regulatory factor) is an ER transmembrane protein known to be bind to STIM1, keeping it in an inactive state. It primarily localizes to the ER membrane and possibly to the plasma membrane (Palty et al., 2012, Albarran et al., 2017). SARAF was discovered in a functional expression screen to identify proteins that affect mitochondrial calcium homeostasis (Palty et al., 2012). HEK293 cells transfected with SARAF cDNA showed lower baseline cytosolic, ER, and mitochondrial calcium levels. In addition, siRNA-based knockdown of SARAF resulted in an increase of basal cytosolic and ER calcium levels. This suggested a role for SARAF in cellular calcium homeostasis. Further experiments conducted using SARAF in conjunction with Orai1 and STIM1 led to deciphering its role in calcium entry. Electrophysiology recordings of Jurkat T cells and HEK293 cells showed SARAF regulates SOCE, specifically the slow calcium-dependent inactivation of CRAC channels (Palty et al., 2012). Subsequent experiments conducted in SH-SY5Y and NG115-401L cell lines also showed that SARAF is expressed in the plasma membrane where it interacts with and negatively regulates ARC channels (Albarran et al., 2016c). In addition, SARAF can also inhibit calcium entry mediated by TRPC1 channels (Albarran et al., 2016b).

How SARAF affects STIM1 structure and function is still not completely understood. Several reports indicate that SARAF interacts with STIM1 in resting conditions to keep STIM1 in its inactive state (Jha et al., 2013, Albarran et al., 2016a). Co-immunoprecipitation studies have shown that SARAF binds to STIM1, and this binding is mediated by the C-terminus of SARAF. The TM and ER-luminal domains are not required for this interaction (Palty et al., 2012). Interestingly, constitutively active STIM1 which has four glutamates mutated to alanine in its SOAR domain does not bind to SARAF and has deficits in slow calcium-dependent inactivation (Jha et al., 2013). The physical interaction between STIM1 and SARAF has been confirmed by multiple reports (Jha et al., 2013, Albarran et al., 2016a, Lopez et al., 2019, Albarran et al., 2017). In addition, SARAF has been shown to bind Orai1 and EFHB (EF hand containing family member B) (Maleth et al., 2014, Albarran et al., 2018, Albarran et al., 2016a, Jha et al., 2013, Lopez et al., 2019).

The nature of SARAF binding to STIM1 and Orai1 is interesting, especially when put into the context of SOCE. SARAF binds to STIM1 in resting conditions to keep it from spontaneously activating. Upon store depletion, co-immunoprecipitation experiments have shown that SARAF immediately dissociates from STIM1 within the first minute, but then they re-associate rapidly (Maleth et al., 2014, Albarran et al., 2018, Albarran et al., 2016a). It is possible that SARAF binds to Orai1 after store depletion, not STIM1. The exact sequence of events by which SARAF binds/releases STIM1 and Orai1 during SOCE is difficult to determine using existing techniques. Experiments conducted using fragments of these proteins show that SARAF interacts with STIM1 at the CC1 and the C-terminal inhibitory domain (CTID) of STIM1 (Jha et al., 2013). TIRF microscopy using STIM1-mCherry and SARAF-GFP show that these two proteins co-localize at the ER:PM junctions upon store depletion using thapsigargin. Co-immunoprecipitation of the SARAF N- and C-termini with STIM1 shows that SARAF binds to the C-terminus of STIM1 (Palty et al., 2012).

Membrane phospholipids play a crucial role in the interaction between SARAF and STIM1/Orai1. Phosphatidylinositol 4,5-bisphosphate (PI(4,5)P$_2$) is a membrane phospholipid that mediates many functions in cellular signaling. It has been shown that the CRAC complex translocates between PI(4,5)P$_2$ poor and rich microdomains during SOCE, and this regulates SOCE (Calloway et al., 2011) . STIM1 interactions with Orai1 and PI(4,5)P$_2$ are both dependent on SARAF (Maleth et al., 2014). In addition, localization of Oria1/STIM1 in PI(4,5)P$_2$ microdomains are maintained by stabilizing proteins such as extended synaptogamin 1 (E-Syt1) and septin4 (Maleth et al., 2014). The dynamic regulation of this complex in membrane microdomains during store depletion has yet to be determined with suitable kinetic resolution. More recently, the SOAR domain of STIM1 has been shown to bind interact with plasma membrane phospholipids. Interactions between lysine residues 382, 384, 385, and 386 in the SOAR domain with plasma membrane phospholipids PI(3,4)P$_2$, PI(3,5)P$_2$, PI(4,5)P$_2$, and PI(3,4,5)P$_2$ are crucial for the stability of ER:PM junctions as well as SOCE (Jha et al., 2013). The Orai1/STIM1 complex localized to the PI(4,5)P2 rich subdomains binds to SARAF and this localization and binding is crucial for SARAF to regulate slow calcium dependent inactivation (Maleth et al., 2014). These complex binding interactions between the SOAR and CTID domains of STIM1 and membrane phospholipids are crucial for the activity of SARAF.

How SARAF regulates slow calcium-dependent inactivation is still unknown. Multiple hypotheses have been proposed to explain this phenomenon. One of them is that an additional calcium-binding protein mediates SARAF activation in slow calcium-dependent inactivation (Zhang et al., 2020). Calcium binding proteins have been known to modulate SOCE, and some of them have been discussed earlier in the review such as EFHB and calmodulin (Mullins et al., 2009). Other hypotheses include modulation by the calcium sensing abilities of STIM1 and Orai1.

**Selectivity and permeation of CRAC channels**

One of the most distinguishing properties of CRAC channels is their very high selectivity toward calcium over other cations. CRAC channels show 1,000 times higher selectivity over sodium. Mutagenesis of the residues lining the channel pore have provided insights into the mechanisms mediating the remarkable selectivity of Orai1 channels toward calcium. Alanine substitutions of E106 in TM1 region, D110, D112, and D114 residues in the TM1-TM2 loop, and E190 in the TM2-TM3 loop all result in a loss of calcium selectivity and increased cesium permeating. In addition, these mutations also diminish calcium-dependent inactivation of these channels. Cesium permeability reveals that the loss of selectivity is likely due to an increase in pore diameter (Yamashita et al., 2007). The E190Q mutation in the TM2-TM3 loop also resulted in loss of calcium selectivity (Vig et al., 2006a) (Prakriya et al., 2006). Also, alanine substitution of the D180 residue changed these channels from calcium selective and inward rectifying channels to sodium/cesium selective and outward rectifying channels (Yeromin et al., 2006). These observations highlight the selectivity filter in the Orai1 channels is formed by the conserved acidic residues lining the outer pore of the channel.

Cysteine scanning experiments showed that the channel pore diameter is largely responsible for the low conductance and high calcium selectivity. Lack of cysteine reactivity suggests that the E190 residue doesn't form the channel pore. Molecular dynamics simulation suggested that mutating this residue to glutamate (E190Q) decreases the pore hydration of the channel (Alavizargar et al., 2018). In addition, mutating this residue to a cysteine (E190C) decreased the calcium selectivity of Orai1 channel (Yeung et al., 2018). It is possible this residue helps in maintaining the pore size by regulating hydration of the ion conduction pathway. Finally, cysteine substitution of aspartate residues at 110, 112, 114 in the TM1-TM2 loop did not change calcium selectivity of these channels (McNally et al., 2009, Zhou et al., 2010a). These results show that TM1 lines the channel pore and a E106 residue (E178 dOrai1) from each monomer forms the selectivity filter in the extracellular side of Orai1 channel.

The combination of electrophysiological analysis and high-resolution structures have resolved many questions regarding Orai structure/function. Orai is unique with regards to subunit composition, ion permeation properties, and the kinetics of gating compared to other calcium channels. Indeed, they have very little sequence homology to other calcium channel families such as voltage-gated calcium channels and ligand-gated channels. Orai channels may have evolved from the very large and ubiquitous Cation Diffusion Facilitator (CDF) carrier family (Matias et al., 2010).

It has been shown that store depletion leads to the relatively slow activation of Orai1 channels (seconds to tens of seconds) (Prakriya and Lewis, 2006). A non-linear gating mechanism has been proposed to regulate CRAC channel gating, where these channels exhibit a 'modal gating' mechanism where the channels alternate between silent and high open probability (Yamashita and Prakriya, 2014). Modal gating mechanism was based on an observation that 2-APB, a widely used as $I_{CRAC}$ inhibitor, elicits strong SOCE at low concentrations and inhibition at high concentrations. This hypothesis proposed that STIM1 binding to the Orai1 channel increases the time spent by these channels in the high open probability state. And because the frequency of these transitions is lower, calcium entry is enhanced (Yamashita and Prakriya, 2014).

Taken together, how calcium depletion in the ER leads to activation of STIM1 shows a complicated picture of multiple events happening in a precise sequential order mediated by several domains of STIM1. First, the binding of calcium ions to the EF-hand keeps the EF-SAM in a compact state which helps interactions between CC1 and CAD regions of STIM1 in the cytosolic side. An inhibitory clamp formed by intramolecular interactions between CC1/2/3 regions of STIM1 helps keep it in inactive state (Yang et al., 2012, Yu et al., 2013, Xu et al., 2006). Store depletion and calcium dissociation from EF-hands leads to a conformational change that releases this clamp and formation of a coiled-coil dimer between CC1 domains. Experiments done using isolated cytosolic STIM1 fragments show that in resting state, these cytosolic fragments are tightly bound, which are extended upon binding to Orai1 (Muik et al., 2011). Mutations in the CC1 regions that cause spontaneous activation of STIM1 also show a similar phenotype (Fahrner et al., 2014). Finally, several leucine residues (L251, L261, L419, and L416) in the CAD region play a crucial role in extension of STIM1 toward plasma membrane for STIM1 binding to Orai1 (Ma et al., 2015, Fahrner et al., 2014, Zhou et al., 2013). This intra-and-intermolecular choreography of STIM1 proteins upon store depletion controls SOCE. An overview of important binding events between STIM1 and plasma membrane lipids is shown in **Figure 2**. These amino acid residues of STIM1 that interact with plasma membrane are also highlighted using an AlphaFold model of the compact and extended conformations of STIM1 (Jumper et al., 2021).

**Orai1/STIM1 clustering/puncta**

Orai1 in the plasma membrane and STIM1 in the ER membrane are diffusely distributed in cells with replete ER calcium stores. Upon store depletion, Orai1-STIM1 complexes are recruited to ER:PM junctions where they form CRAC channel clusters to promote efficient calcium entry upon store depletion (Liou et al., 2007). These clusters were denoted as "puncta" upon observation under fluorescence microscopy of tagged-STIM1 (Xu et al., 2006, Zhang et al., 2005, Wu et al., 2006, Baba et al., 2006, Liou et al., 2005, Luik et al., 2006). These ER:PM junctions are regions within the cell where PM and ER are held in close apposition (~10-20 nm) (Orci et al., 2009, Wu et al., 2006). Upon store depletion, STIM1 accumulates near the thin cortical tubules of the ER (Orci et al., 2009). An interesting observation in CRAC puncta formation is the longevity of these puncta. Depending on the downstream effect of calcium influx, the puncta can be active for a few minutes or up to an hour. How these puncta can stay active to promote calcium entry for this duration is still not completely understood. Even more compelling is the ability of STIM1 to translocate to the sites of already formed puncta repeatedly upon subsequent store depletion (Smyth et al., 2008). A multitude of factors regulate this behavior to increase the number of cortical ER tubules near the plasma membrane. One mechanism is a secretion-like coupling model which includes redistribution of F-actin into cortical layers (Patterson et al., 1999). STIM1 has been known to interact with microtube attachment protein EB1 and maintain ER tubule length (Grigoriev et al., 2008). STIM1 also binds to plasma membrane using its polybasic domain, which might strengthen the STIM1 localization in the puncta (Liou et al., 2007). The SOAR region of STIM1 also has a cholesterol binding domain which has been shown to bind to cholesterol rich regions in the plasma membrane (Pacheco et al., 2016). Several additional proteins like synaptogamins and septins maintain the integrity of these ER:PM junctions (Sharma et al., 2013, Chang et al., 2013). Cytosolic calcium elevation leads to increased translocation of E-Syt1 to ER:PM junctions, which subsequently recruits the phosphatidylinositol transfer protein (PITP) Nir2 to these junctions to strengthen ER:PM junction stability (Chang et al., 2013). E-Syt2 and E-Syt3 also regulate the ER:PM junctions, in a calcium-independent manner (Giordano et al., 2013). Interestingly, siRNA-mediated knockdown of E-Syt proteins decreases the number of ER:PM contact sites but does not affect SOCE (Giordano et al., 2013). This explains the possibility that E-Syt proteins maintain the stability of ER:PM junctions with no specific effect on Orai1 or STIM1. Another protein that has shown a role in regulating SOCE as well as long-term maintenance of ER:PM junctions is an ER transmembrane protein TMEM110, also known as junctate (Quintana et al., 2015). Junctate is a calcium binding protein in the ER membrane that forms supramolecular complexes with $IP_3$ receptors and TRPC3 calcium channels. Junctate has an EF-hand domain in its ER luminal side, which is required for CRAC channel activation independently of store depletion (Srikanth et al., 2012, Treves et al., 2004). In addition to these proteins mentioned above, many other proteins play a crucial role in maintenance of stable ER:PM junctions that help CRAC channel formation and function (Ivanova and Atakpa-Adaji, 2023).

## Models of CRAC channel assembly

### ATP-dependent puncta formation

Based on the observation in live cells that STIM1 translocates to the plasma membrane from the ER membrane and an earlier hypothesis that STIM1 is transported to PM via a secretory pathway (Hauser and Tsien, 2007), puncta formation was hypothesized to be ATP-dependent. Mitochondria are known to accumulate near the ER:PM junctions and regulate calcium homeostasis (Quintana et al., 2011, Park et al., 2001). In addition, depletion of intracellular ATP leads to decreased calcium entry in rat lymphocytes (Marriott and Mason, 1995). ATP depletion can also lead to translocation of STIM1 to puncta. However, this required depletion of $PI(4,5)P_2$ in addition to ATP depletion (Chvanov et al., 2008).

### Microtube-associated STIM1 translocation

As discussed earlier in the review, STIM1 binds to the EB1 protein that is known to modulate microtubule growth. In B cells, treatment with the anti-mitotic agent nocodazole, which inhibits polymerization of microtubules, did not show the puncta formation observed in untreated cells (Baba et al., 2006). Pull down assay with EB1, EB2, and EB3 proteins showed STIM1 binds to EB1. This was also confirmed using immunocytochemistry and co-immunoprecipitation experiments (Grigoriev et al., 2008). In addition, treatment of cells with ML-9, a myosin light chain kinase inhibitor, led to reversal of CRAC puncta as well inhibition of SOCE (Smyth et al., 2008). These observations led to a hypothesis that STIM1 traffics to the ER:PM junctions by microtubule-assisted transport mediated by its binding to EB1. However, this model shows that STIM1 and Orai1 proteins are confined at the junctions after store depletion but does not offer any explanation to why this happens.

**Phosphatidylinositol-mediated membrane sorting**

Early experiments conducted on STIM proteins led to discovery of a C-terminal polybasic domain that interact with PM phospholipids, such as $PI(4,5)P_2$ (Ercan et al., 2009, Calloway et al., 2011). This polybasic domain was also found to mediate the inward rectification of SOCE currents (Yuan et al., 2009). HeLa cells treated with wortmannin and LY294002, inhibitors of phosphatidyl inositol 3-kinase (PI3K) and PI4K respectively, led to inhibition of STIM1 puncta formation as well as SOCE (Walsh et al., 2009). In addition, the binding between STIM1 and Orai1 was differentially modulated by the levels of $PI(4,5)P_2$ (Calloway et al., 2011). Decreased $PI(4,5)P_2$ concentrations resulted in reduced thapsigargin-mediated Orai1-STIM1 binding in the PM liquid-ordered phase/rafts, and this affinity was reversed in membrane disordered regions (Walsh et al., 2009). This led a hypothesis that membrane phospholipids play a role in CRAC channel formation and SOCE. However, depletion of phosphoinositides in cells overexpressing Orai1 did not affect either STIM1 puncta formation or SOCE (Walsh et al., 2009). In addition, the experiments showing a role for phosphoinositides was shown using inhibitors which do not discriminate between direct effects on Orai1/STIM versus general effects due to severely perturbing PM phospholipid composition.

**$IP_3$-mediated calcium depletion in ER lumen**

Before the discovery of Orai1 and STIM1 as proteins forming CRAC channels, experiments performed on TRP3 transfected HEK293 cells using plasma membrane patches by treatment with $IP_3$. Based on this observation, they proposed a hypothesis where the released $IP_3$ binds to the $IP_3$Rs and activates them, which in turn regulate SOCE and CRAC channels. However, they do not show physical binding between $IP_3$Rs and TRP3 proteins (Kiselyov et al., 1998). This physical binding between the $IP_3$Rs and TRP3 was shown later using coimmunoprecipitation of tagged constructs of $IP_3$R and TRP3 (Boulay et al., 1999). They also show the role of this binding in the regulation of SOCE. In subsequent studies, it was also shown that TRP3 forms store-operated cation channels dependent and independently with $IP_3$Rs using DT40 WT and $IP_3$R knockout cells with rescue expression of TRP3 (Vazquez et al., 2001). However, the mechanism behind this modulation of SOCE by $IP_3$Rs remained unresolved. Recently, it was shown that STIM1 proteins interact with $IP_3$Rs in the ER membrane (Beliveau et al., 2014). Independently, others have shown that overexpression of $IP_3$Rs leads to increased ER calcium depletion, larger CRAC puncta, and higher SOCE (Sampieri et al., 2018). In addition, using confocal microscopy, they also show $IP_3$Rs are recruited to CRAC puncta upon activation of $IP_3$Rs using bradykinin. Based on these observations, they proposed a mechanism for SOCE where recruitment of $IP_3$Rs to CRAC puncta leads to the generation of localized calcium-free microenvironment in the luminal side of ER, which helps in activation of STIM1 (Sampieri et al., 2018).

**Diffusion-trap model**

The diffusion-trap model postulates that STIM1 undergoes a conformational change that exposes its C-terminal domains toward plasma membrane, where it binds to membrane phospholipids and stochastically binds to Orai1 channels laterally diffusing in the PM (Hoover and Lewis, 2011, Wu et al., 2014). In agreement with this hypothesis, super-resolution microscopy experiments conducted using tagged Orai1 and STIM1 constructs showed that these proteins diffuse randomly in resting conditions. Upon store depletion, these proteins slow down at distinct ER:PM junctions allowing them to accumulate and form puncta (Wu et al., 2014). Additionally, single particle tracking analysis shows single STIM1 and Orai1 particles diffusing freely before getting trapped at the junctions. Deletion of the polybasic domain in the C-terminus of STIM1 showed altered puncta formation after store depletion, suggesting a direct role for this domain. These experiments also demonstrated that STIM1 and Orai1 particles have a long half-life in puncta once trapped (Qin et al., 2020, Wu et al., 2014).

The diffusion-trap model fails to explain some aspects of puncta formation and SOCE. The polybasic domain was hypothesized to act as a trap to attract Orai1 to STIM1. However, a mutant STIM1 devoid of this polybasic domain is capable of binding with Orai1. Orai1 binding by itself can also trap STIM1 within ER:PM junctions which raises the possibility that Orai1 may be trapping STIM1 (Walsh et al., 2009). Interestingly, we have found that STIM1 can form puncta in the absence of Orai1, but Orai1 cannot form puncta in the absence of STIM1 (West et al., 2022b, Kodakandla et al., 2022) . The strong binding between Orai1 and STIM1 in the puncta could also mean there is an equivalent amount of protein particles enter and leave the puncta, thereby maintaining a dynamic equilibrium at these ER:PM junctions (Wu et al., 2014). Finally, single particle tracking and polydispersity analyses conducted on Orai1 and STIM1 proteins show that the mobility of these proteins decreases after store depletion, but these proteins are also confined in compartmentalized membrane regions before and after store depletion (Qin et al., 2020). Based on these observations, the binding of Orai1 and STIM1 upon store depletion appears to be more complicated than a simple diffusion-trap model.

**S-Acylation as a regulator of Orai1/STIM1 assembly**

S-acylation is a reversible addition of lipid moieties to cysteine residues of target proteins that affects protein stability, function, conformation, and trafficking between compartments within a cell (Chamberlain and Shipston, 2015, Chen et al., 2021). S-acylation is a post translational modification that is mediated by specific group of enzymes known as protein acyltransferases (PATs). PATs are also known as DHHC enzymes owing to the conserved aspartate-histidine-histidine-cysteine motif in their active site that carries out this reaction (Chamberlain and Shipston, 2015). The different classes of DHHC enzymes distinguish among themselves by their selectivity toward substrates, lipid preferences, and mechanisms that activate and inactivate them (Chen et al., 2021). Deficiencies in DHHC enzyme functions leads to a range of diseases ranging from cancers such as adenocarcinomas, lung cancer, bladder cancer, and breast cancer, to diseases that affect neurological functions such as epilepsy and schizophrenia, glioblastoma, and X-linked intellectual disability (Chen et al., 2021, Essandoh et al., 2020, Fraser et al., 2020, Ladygina et al., 2011, Resh, 2016). The subsequent process of removal of lipid moieties from protein substrates, known as deacylation, is mediated by another set of enzymes known as acyl protein thioesterases (APTs) (Duncan and Gilman, 1998, Lin and Conibear, 2015). These enzymes act in concert to dynamically regulate protein function in a stimulus-dependent manner. How most DHHC and APT enzymes are activated and inactivated is still unclear.

DHHC enzymes catalyze the addition of lipid moieties to proteins through a two-step mechanism, where the cysteine residue in the active site of the enzyme receives the lipid group and then transfers it to the cysteine residues of target proteins (Jennings and Linder, 2012). This first step, known as auto-acylation, is followed by the transfer of acyl group to target proteins. The DHHC enzymatic reaction is sometimes referred to as a "ping-pong" reaction. However, this may be a misnomer. Ping-pong mechanisms (also known as double-displacement reactions) always involve two substrates and two products, with the binding of the second substrate dependent upon the successful completion of the first reaction (Kullmann, 1984, Ulusu, 2015). The reaction mechanism for DHHC enzymes is highlighted in **Figure 3**. For the DHHC reaction to be considered ping-pong, we would have to accept that CoA is product #1, and that substrate

#2 binding (the protein be S-acylated) depends upon autoacylation of the DHHC enzyme. We would argue that the autoacylated DHHC enzyme is an intermediate, not an enzyme newly capable of binding substrate #2. Regardless, the ultimate result is the S-acylation of the target protein. Motivated by the finding that the protein kinase Lck is dynamically S-acylated in T cells during Fas signaling, we investigated the role(s) of this modification in SOCE pathways (Akimzhanov and Boehning, 2015).

Orai1 and STIM1 accumulate in lipid raft domains upon depletion of intracellular calcium stores. Depletion of PM cholesterol with methyl-beta-cyclodextrin impairs SOCE, implying a role for lipid rafts in SOCE (Jardin et al., 2008b, Dionisio et al., 2011b). As mentioned above, our group found that signaling through the Fas death receptor in T cells requires dynamic S-acylation of the kinase Lck, leading to translocation to rafts where it activates PLC-γ1-mediated IP$_3$ production (Wozniak et al., 2006). Subsequently, we found that many TCR components such as Lck, Fyn, ZAP70, and PLCγ1 undergo dynamic S-acylation upon T cell activation (West et al., 2022a, Akimzhanov and Boehning, 2015, Akimzhanov et al., 2010). This S-acylation was required the calcium/calmodulin-dependent activation of the protein acyltransferase DHHC21. We next investigated T cell function in the mutant mouse *depilated*, which carry a functionally deficient DHHC21 with an in-frame deletion of phenylalanine 233 (F233) eliminated calmodulin binding. *Depilated* mice have severe deficits in T cell differentiation, including differentiation into Th1, Th2, and Th17 lineages (Bieerkehazhi et al., 2022). TCR signaling in *depilated* mice is severely disrupted due to reduced activation of ZAP70, Lck, PLCγ1, JNK, ERK1/2, and p38 in response to TCR stimulation (Bieerkehazhi et al., 2022, Fan et al., 2020). As Orai1 and STIM1 are essential for TCR signaling, we hypothesized that Orai1 and STIM1 undergo S-acylation to regulate CRAC channel formation and function in T cells.

We found that treating Jurkat T cells with anti-CD3 or HEK293 cells with thapsigargin leads to stimulus-dependent S-acylation of Orai1 and STIM1. The kinetics of S-acylation Orai1 and STIM1 are very rapid consistent with the assembly of puncta and channel activation. The cysteine-mutant versions of Orai1 and STIM1 that are incapable of undergoing S-acylation have significantly reduced puncta formation and SOCE, indicating a direct role for S-acylation in CRAC channel assembly (West et al., 2022b, Kodakandla et al., 2022).

One interesting observation between the cysteine mutants of Orai1 and STIM1 is that the Orai1 cysteine mutant (C143S) was an almost complete loss of function phenotype, whereas mutant STIM1 (C437S) retained some residual activity. The residue where STIM1 undergoes S-acylation is at the proximal end of its SOAR domain. As explained earlier in the review, the SOAR domain of STIM1 is crucial in the tethering of STIM1 to plasma membrane. The C-terminus of STIM1 also has the polybasic membrane binding domain. We hypothesize the S-acylation of STIM1 stabilizes the C-terminus in the plasma membrane. Due to redundant membrane-binding domains in the C-terminus, we postulate that the loss of S-acylation would only partially reduce the affinity of the C-terminus for the membrane, thus possibly explaining our partial loss-of-function phenotype (**Figure 2**) (Pacheco et al., 2016). Our model is that a plasma membrane resident DHHC enzyme, probably DHHC21, S-acylates STIM1 and anchors its C-terminus in the plasma membrane, thereby stabilizing the CRAC complex at these ER:PM junctions to facilitate SOCE (Kodakandla et al., 2022, West et al., 2022b). An overview of the models suggested in this section is presented in **Figure 4**.

At the same time as our research into S-acylation of Orai1 and STIM1, the Demaurex group also showed that Orai1 undergoes S-acylation (Carreras-Sureda et al., 2021). It was found that S-acylation of Orai1 is critical for its recruitment to the immunological synapse as well as TCR-mediated calcium entry in Jurkat T cells. Using TIRF and confocal imaging techniques, they showed that S-acylation of Orai1 is critical for channel clustering as well as their trafficking to the lipid rafts. Finally, using overexpression analysis, they show that DHHC20 is the enzyme that mediates S-acylation of Orai1 in the plasma membrane, at least under resting conditions.

We hypothesize that DHHC20 may mediate Orai1 S-acylation under resting conditions, whereas store depletion leads to calcium/calmodulin-dependent activation of DHHC21 leading to increased Orai1/STIM1 S-acylation and active partitioning to ER:PM junctions. This would be consistent with our data in *depilated* DHHC21 mutant mice showing altered T cell calcium signaling in response TCR activation (Tewari et al., 2021, Fan et al., 2020, Bieerkehazhi et al., 2022).

**Unanswered questions and further directions**

Our hypothesis that S-acylation plays a role in store-operated calcium entry was derived from the inability of current models to explain the specific characteristics that make this process unique. As explained earlier in the review, Orai1 and STIM1 proteins show distinct features upon store-depletion, such as decreased mobility in the membrane, targeting to membrane subdomains, and dynamic nature of CRAC puncta, among other observations. We based our hypothesis that S-acylation, being a process that is controlled by a set of enzymes, gives a dynamic control to Orai1 and STIM1 function.

Of the many topics not answered in our studies, we want to highlight a few that are more intriguing and necessitate scrutiny. We have discussed the role of DHHC enzymes in S-acylating CRAC channel components, including DHHC20 and DHHC21. But the enzymes that deacylated these proteins have not been studied yet. As explained earlier, S-acylation is reversible. Acyl protein thioesterases (APT) are responsible for removing the lipid moiety from S-acylated proteins. APT enzymes have been thought to be constitutively active. For example, radiolabeling studies using tritiated palmitate have shown than palmitate incorporation has a half-life of 20 minutes whereas depalmitoylation is 10-20 times faster (Magee et al., 1987, Rocks et al., 2005). Another difference between S-acylation and deacylation is the localization of proteins that undergo this process. Some reports suggest S-acylation is restricted to compartments where the specific DHHC enzyme resides, such the Golgi apparatus, while deacylation can happen throughout the cell (Rocks et al., 2010). However, the specific mechanisms how DHHC and APT enzymes coordinate these steps has not been explored. Interestingly, APT enzymes themselves are targets for S-acylation, indicating the DHHC enzymes regulate APT enzyme localization and function (Abrami et al., 2021). Nothing is known about the regulation of Orai1/STIM1 by APT enzymes other than the S-acylation of both proteins is transient after store depletion, suggesting a direct for APT enzymes in puncta disassembly (Kodakandla et al., 2022, West et al., 2022b) .

How S-acylation drives the movement of these proteins to subdomains in the plasma membrane is also interesting. Addition of lipid moieties (long chain lipids) changes the hydrophobicity of the proteins. The result of this addition is an increased affinity toward phospholipids, cholesterol, or other moieties in membrane subdomains (Greaves and Chamberlain, 2007). For example, addition of saturated lipids can result in translocation to cholesterol and sphingolipid-enriched membrane subdomains such as lipid rafts (Pani and Singh, 2009, Resh, 2016). Indeed, previous work by our group has shown that S-acylation is critical for the assembly of the TCR complex in lipid rafts (Tewari et al., 2021, Akimzhanov et al., 2010, Akimzhanov and Boehning, 2015, Fan et al., 2020).This might also help us understand the differences observed between cysteine mutants of Orai1 and STIM1, which show different calcium entry patters upon store-depletion.

Another interesting point of emphasis which we failed to make is the specific lipid moiety that is added to target proteins, and how the lipid moiety affects the protein behavior. The lipid moiety could range from saturated palmitate (C16:0, 74%) to monounsaturated oleate (C18:1, 4%) or saturated stearate (C16:0, 22%) among others (Towler and Glaser, 1986). The specific type of lipid moiety attached to Orai1 and STIM1 will help answer some questions regarding the function of the S-acylated residue. For example, we hypothesized S-acylation of Orai1 helps it recruit to cholesterol-rich lipid rafts, whereas STIM1 S-acylation anchors the SOAR domain of STIM1 in the plasma membrane to facilitate its interaction with Orai1 in lipid rafts. Palmitic acid has long known to have higher affinity toward cholesterol compared to other phospholipids (Melkonian et al., 1999, Levental et al., 2010). So, we can test if Orai1 S-acylation leads to addition of a palmitate group using metabolic labeling. On the other hand, stearic acid is known to recruit to plasma membrane domains that have low levels of phosphatidylinositol-4,5-bisphosphate ($PIP_2$) (Laquel et al., 2022). Some reports have suggested the CRAC channels translocate to these domains upon store-depletion for propagating SOCE currents (Maleth et al., 2014). Stearic acid conjugation to STIM1 may lead to differential association with plasma membrane domains depleted of $PIP_2$. Delineating these hypotheses is crucial for understanding the role of S-acylation in SOCE.

The S-acylation of STIM1 may stabilize the C-terminus of the active conformation at the plasma membrane. In addition to S-acylation, two other critical binding events happen between the C-terminus of STIM1 and the plasma membrane. In particular, STIM1 binds to membrane lipids on the inner leaflet via the polybasic domain and the cholesterol-binding domain (Figure 2; (Yuan et al., 2009, Pacheco et al., 2016). This functional redundancy of membrane-associated features in the C-terminus of STIM1 may explain the partial loss of function in the cysteine-mutant STIM1. Future work can test this hypothesis using different STIM1 mutants that lack the polybasic or CBD domains alone and in combination with the S-acylation mutant.

Finally, how the DHHC enzymes that S-acylate Orai1 and STIM1 are regulated is a question that remains unanswered. There is surprisingly little information about how these enzymes are regulated. Previously, we have shown that DHHC5, an enzyme the S-acylates many signaling proteins involved in beta adrenergic receptor function, is itself regulated by S-acylation of several cysteines in the C-terminal tail. S-acylation of DHHC5 in its C-terminal tail upon receptor stimulation increases its stability in the plasma membrane where it can S-acylate its target proteins (Chen et al., 2020). In addition, the DHHC enzymes regulate each other, where one

DHHC enzyme can S-acylate another DHHC enzyme, which in turn affects its localization, function, or stability. For example, DHHC16 S-acylated DHHC6, which is an ER membrane protein that S-acylates many ER proteins such as calnexin, E3 ligase gp78, and IP$_3$ receptor (Abrami et al., 2017). We previously showed that DHHC21 is a calcium-calmodulin regulated enzyme S-acylates many T cell proteins in vitro and in vivo. Future studies will determine if DHHC21 is also the key regulator of Orai1/STIM1 S-acylation. Ultimately, the regulation of CRAC channel formation and SOCE by S-acylation adds a regulatory step for channel assembly and disassembly during store depletion. This allows for the fine tuning of the spatiotemporal aspects of calcium signaling in cells expressing Orai1/STIM1.

**List of Figures:**

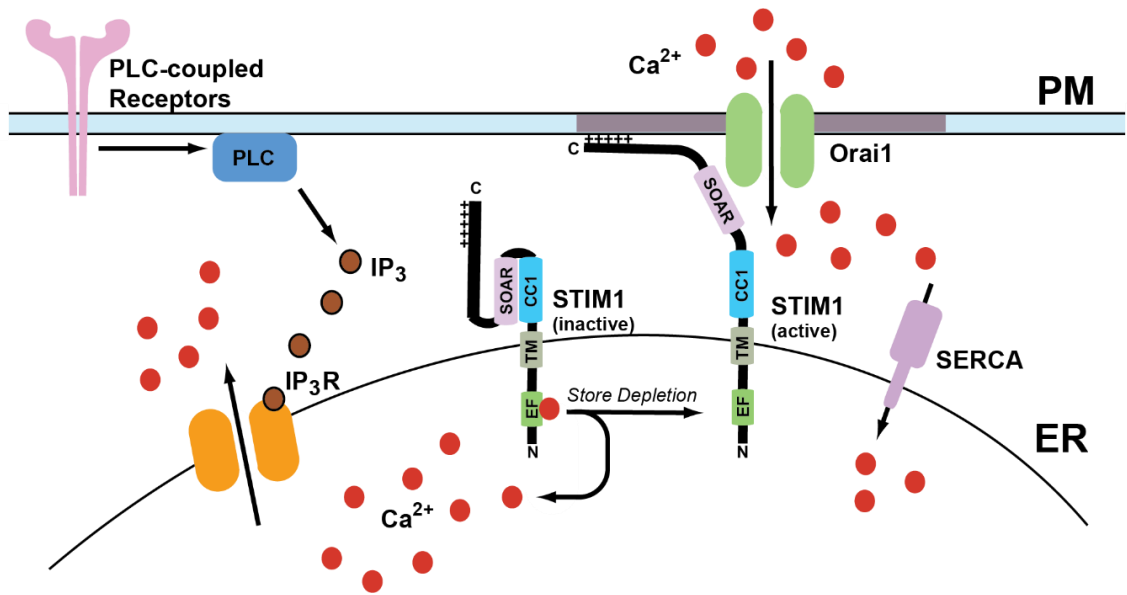

**Figure 1: General overview of store-operated calcium entry (SOCE).** Agonist stimulation through the PLC-coupled receptors leads to $IP_3$-mediated calcium release which leads to ER calcium store depletion. As a result, calcium dissociates from the EF-hand of STIM1 and leads to its activation. This leads to a conformation change in STIM1 which extends its C-terminus toward plasma membrane where it binds to Orai1 channels and form calcium-release activated calcium (CRAC) channels. CRAC channels promote calcium entry from extracellular milieu into the cells. One subunit of STIM1 dimer is shown here for simplicity.

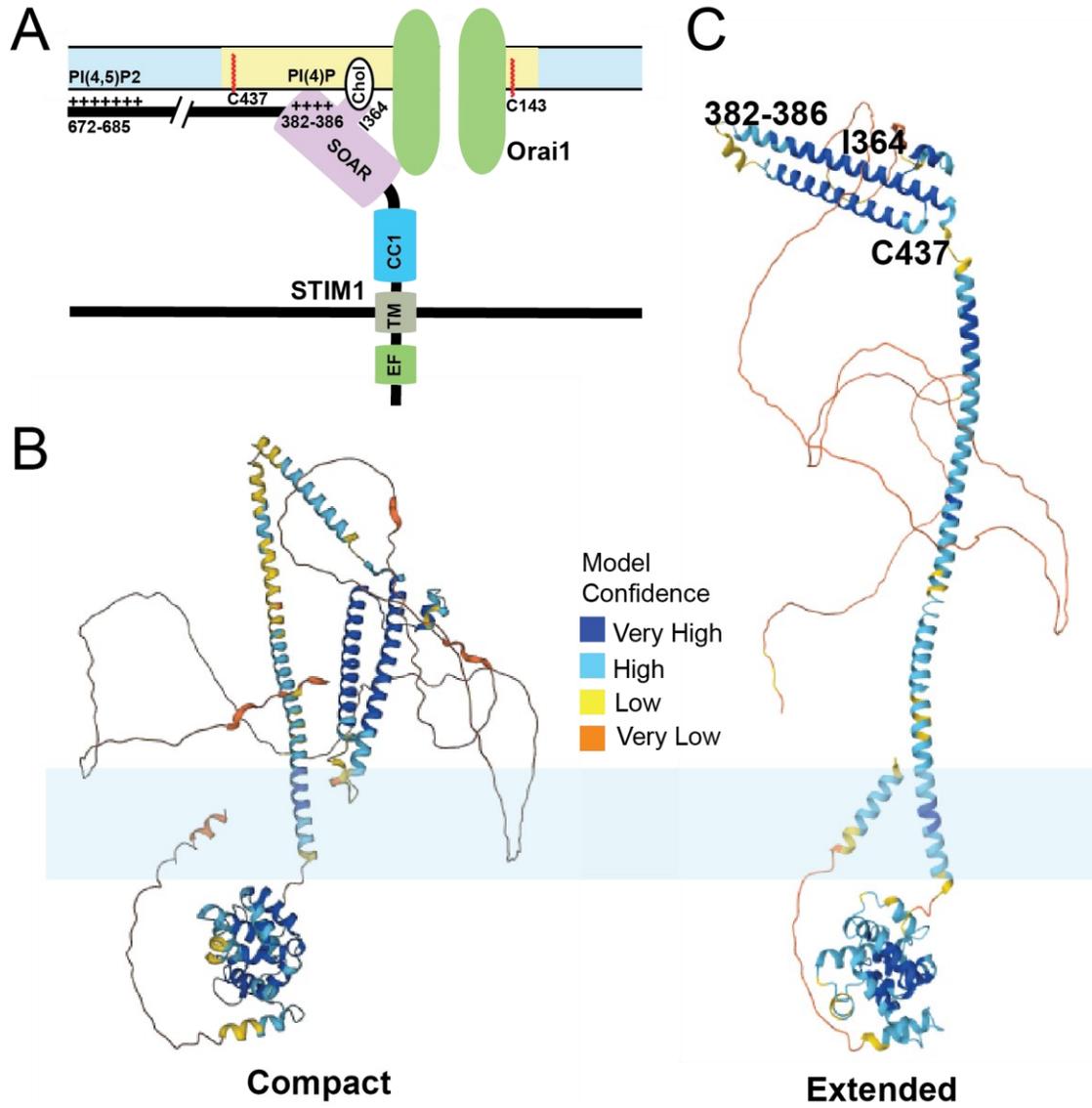

**Figure 2: STIM1-lipid binding at the plasma membrane:** A) The CRAC channels are stabilized at the ER:PM junctions by a multitude of bindings between different domains of Orai1 and STIM1 with plasma membrane lipids upon store depletion. Orai1 undergoes S-acylation at its C143 residue which shuttles the channels to lipid rafts. The SOAR domain of STIM1 binds to the N-terminus region of Orai1 and leads to activation of Orai1 channels. The cholesterol binding domain (CBD) in the SOAR domain also binds to the cholesterol rich phospholipids in the plasma membrane, which is mediated by I364 residue. The polybasic domain of STIM1 binds to the PI(4,5)P2 phospholipids in the plasma membrane which is mediated by the positively charged amino acids in the C-terminal tail of STIM1. Finally, STIM1 undergoes S-acylation at its C437 residue which is crucial for SOCE. One subunit of STIM1 dimer is shown here for simplicity. B & C) The AlphaFold structure predictions of the compact (presumably inactive) and extended (presumably active) conformations of STIM1 are shown. The residues on STIM1 that interact with the plasma membrane as well as C437 that undergoes S-acylation are highlighted in Panel C. In B and C, the lipid bilayer is represented in light blue. The compact model is the AlphaFold prediction of Uniprot #V5J3L2. The extended model is the AlphaFold prediction of Uniprot #Q13586).

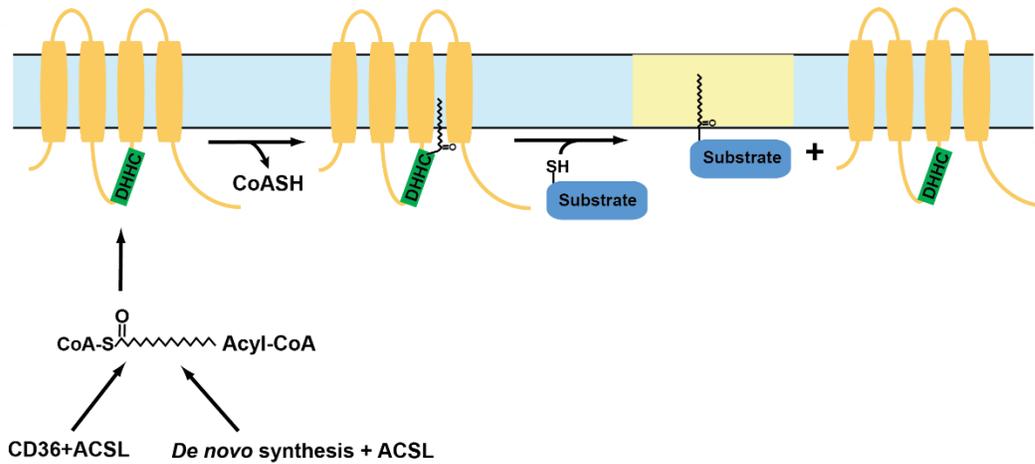

**Figure 3: Mechanism of DHHC enzyme catalysis.** DHHC enzymes S-acylate their substrates by a two-step mechanism which involves autoacylation followed by transfer of the acyl group to target protein cysteine residues (substrates). In this cartoon, lipid rafts are indicated in yellow. ACSL = Acyl-CoA synthetase long-chain family members.

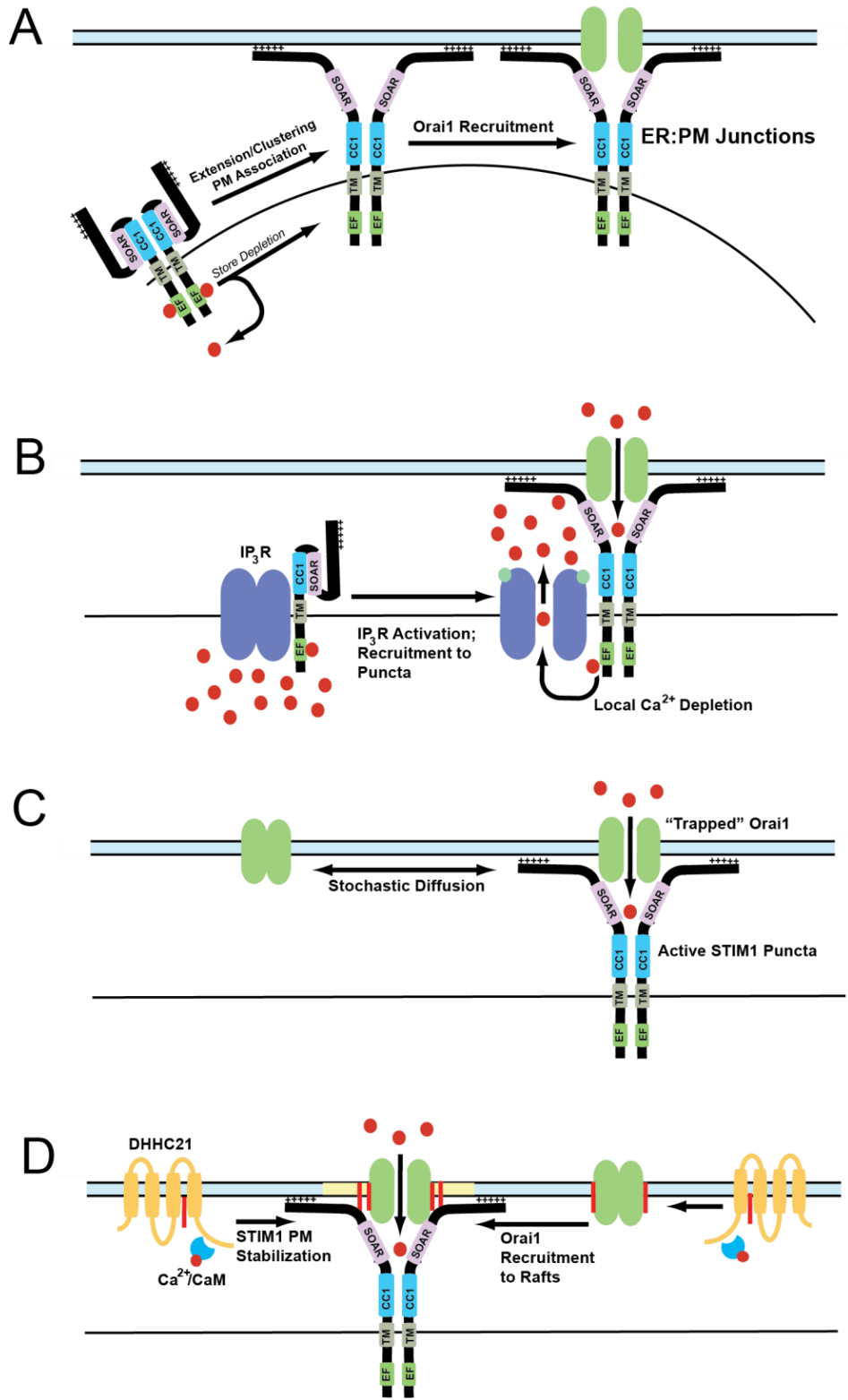

**Figure 4: Models proposed to explain CRAC channel assembly in SOCE.** A) Overview of CRAC puncta formation after ER calcium store-depletion. B) IP$_3$R recruitment to puncta with STIM1 and IP$_3$-mediated local calcium depletion C) The Orai1 diffusion trap model, and D) S-acylation of Orai1 and STIM1 and recruitment to lipid rafts. The three models are not necessarily mutually exclusive. See the text for details.

**Appendix, Abbreviations**

| | |
|---|---|
| acyl-RAC | acyl resin-assisted capture |
| BAPTA | l,2-bis(0-aminophenoxy)ethane-A;A^A',,A''-tetraacetic acid |
| BSA | bovine serum albumin |
| C-terminal | carboxy terminal |
| CaM | calmodulin |
| CRAC | calcium release-activated calcium |
| DHHC | aspartate-histidine-histidine-cysteine |
| DTT | dithiothreitol |
| EDTA | ethylene diamine-tetraacetic acid |
| EGTA | ethylene glycol-bis(P-aminoethyl ether)N,N,N',N'-tetraacetic acid |
| ER | endoplasmic reticulum |
| HA | hydroxylamine |
| HEK293 | human embryonic kidney 293 cell line |
| IP3 | inositol 1,4,5 trisphosphate |
| IP3R | inositol 1,4,5 trisphosphate receptor |
| mRFP | monomeric red fluorescence protein |
| PAT | protein acyltransferase |
| PBS | phosphate buffered saline |
| PIP2 | phosphatidylinositol 4,5 bisphosphate |
| PLC | phospholipase C |
| PM | plasma membrane |
| PMSF | phenyl methyl sulfonyl fluoride |
| SDS-PAGE | sodium dodecyl sulfide polyacrylamide gel electrophoresis |
| SERCA | sarco-endoplasmic reticulum calcium ATPase |
| SOCE | store-operated calcium entry |
| STIM1 | stromal interaction molecule 1 |
| TCR | T-cell receptor |
| TG | thapsigargin |
| TIRF | total internal reflection fluorescence |

**Declarations:**

**Author contributions:**

All authors contributed to writing the manuscript. GK wrote the first draft. GK, AA, and DB edited the draft. GK and DB prepared the figures. GK and DB finalized the manuscript.

**Funding:**

This work was supported by grants from the National Institute of General Medicine and Sciences (NIGMS) of the National Institutes of Health (NIH) under award numbers R01GM130840 (to D. B. and A. M. A.), R01GM115446 (to A. M. A.), and R01GM081685 (to D. B.) and startup funds provided by Cooper Medical School of Rowan University (DB).

**Acknowledgements:** Not Applicable

**Authors' information:**

Goutham Kodakandla[1*], Askar M. Akimzhanov[2], Darren Boehning[1*]

[1] Department of Biomedical Sciences, Cooper Medical School of Rowan University, Camden, NJ, USA, 08103

[2] Department of Biochemistry and Molecular Biology, McGovern Medical School, University of Texas Health Sciences Center at Houston, Houston, Texas, USA, 77030

[*]For correspondence, GK (kodaka78@rowan.edu) or DB (boehning@rowan.edu)

**Conflict of interest:**

The authors declare that the research was conducted in the absence of any commercial or financial relationships that could be considered as potential conflict of interest.